# Sustainability in Software Product Lines: Report on Discussion Panel at SPLC 2014

Ruzanna Chitchyan
Department of Computer Science
University of Leicester
Leicester, UK
rc256@le.ac.uk

Joost Noppen
School of Computing Sciences
University of East Anglia, UK
j.noppen@uea.ac.uk

Iris Groher
Johannes Kepler University Linz
Linz, Austria
iris.groher@jku.at

## ABSTRACT

Sustainability (defined as "the capacity to keep up") encompasses a wide set of aims: ranging from energy efficient software products (environmental sustainability), reduction of software development and maintenance costs (economic sustainability), to employee and end-user wellbeing (social sustainability). In this report we explore the role that sustainability plays in software product line engineering (SPL). The report is based on the "*Sustainability in Software Product Lines*" panel held at SPLC 2014.

## Categories and Subject Descriptors

D.2.8 [**Software**]: Metrics – *performance measurement, process metrics, product metrics.* D.2.9 [**Software**]: Management – *productivity, software quality assurance.* K.4.2 [**Computing Milieux**]: Social issues – *employment.* K.4.2 [**Computing Milieux**]: Organizational Impacts – *automation, computer-supported collaborative work, employment, reengineering*

## 1. INTRODUCTION

Sustainability (defined as "the capacity to keep up") encompasses a wide set of aims: ranging from energy efficient software products (environmental sustainability), reduction of software development and maintenance costs (economic sustainability), to employee and end-user wellbeing (social sustainability). In this report we explore the role that sustainability plays in software product line engineering (SPL).

This report is based on the "*Sustainability in Software Product Lines*" panel held at SPLC 2014. The report brings together ideas discussed by the panellists as well as audience members.

***Panellists*** (ordered by surnames):

- *Dr. Danilo Beuche: CEO at the Pure Systems, Germany*
- *Dr. Paul Clements: Vice President of Customer Success at BigLever Software, USA*
- *Prof. Mark Harman: Professor of Software Engineering at University College London, UK*
- *Prof. Linda Northrop: Chief Scientist and SEI Fellow at the Software Engineering Institute and a Professor at Carnegie Mellon University, USA*
- *Dr. Rick Rabiser: senior researcher at Johannes Kepler University Linz, Austria*

This panel explored state, challenges, and directions in research and practice for Sustainability in Software Product Line Engineering. These are summarised in section 2 of the present report. The panel also discussed the relation and applicability of the principles and commitments of Sustainability Design, as stated in the Karlskrona Manifesto for Sustainability Design [1]. The relevance of these principles and commitments to Software Product Lines Engineering is discussed in section 3 of this report.

## 2. WHAT IS "SUSTAINABILITY" FOR SPL?

### Economic Sustainability

For **SPL** sustainability has, first and foremost, **economic implications**. An **SPL is a business model** adapted by a company in order to succeed with its business objectives, or, in other words, achieve financial profit. Once a company succeeds, it wishes to continue to enjoy the level of its success. Thus, sustainability here is the ability to continue to enjoy financial benefit, that is the sustainability of the business, not the PL per-se. Change, however, undermines economic sustainability of SPL. Change can have a number of origins, such as related to technology, governance, or people. *SPL embraces change by maintaining commonality and managing variability.*

To be sustainable an SPL must be:

1. resilient to change and threats, without much extra re-investment. This implies ability to manage the variability of the technical asset base, variability of context in which the business exists (i.e., its ecosystem and governance), and variability of people.

2. adaptive, i.e., able to support change which is not covered by SPL, but is disruptive.

Moreover, from the economic perspective, sustainability of an SPL-based business is also pre-supposes expected evolution of the business beyond a single PL. This is because a dramatic change often happens at a larger scope than a single PL. Such change (e.g., move to a completely new market or technological solutions) maybe force evolution of a given PL into a new "off-spring" PL, or even change of the business model from SPL to something else.

Thus, when considering economic sustainability with SPL, the goal is not to keep a product line going for as long as possible, as sustainability of a business is not always about sustaining an SPL.

### Social Sustainability

The social sustainability perspective in SPL is particularly relevant for the **processes** that support SPL business model, and so, the SPL product.

From the **process-focused** perspective, in order to ensure sustainable SPL model, the respective SPL-focused culture must be established in the company. The main pressures on the SPL process come from:

1. people mobility, whereby departure of key people from the company leads to loss of knowledge and process disruption. To counter this, it is necessary to document the **core PL process variability** to help sustain PL beyond employment commitments of specific individuals. This point is also relevant to with respect to SPL champions. Often these are innovative individuals who invigorate the SPL practice in a company, but a champion must become a "community of practice" to allow for the SPL to continue in his/her absence.

2. process "workarounds", whereby some quick fixes could be used in SPL, e.g., when the proper process knowledge is missing, which subvert the PL process and product. The workarounds often arises when there is a knowledge/skills loss due to people mobility, or simply due to deadline pressures.

Thus, in order to support process sustainability, the company needs to monitor and measure SPL process and product. Moreover, it is necessary to ensure that the infrastructure for monitoring and measurement itself is sustainable.

With respect to social sustainability, there is misconception that **adoption of the SPL model can lead to job loss** or lessen social sustainability. The SPL practitioners and researchers unanimously agree that the adoption of SPL practices reduces tedium in jobs and capitalizes on the workers experience, but it does not lead to job loss. SPL helps to maintain business revenue and leads to a more diverse ecosystems around it.

Yet, proper use of SPL must be accompanied with automation for such tasks as variability and commonality management, configuration derivation and test specification. Without such essential automation, SPL will be much more "counter-sustainable", as many tasks that are, by design, intended for a tool, will have to be handled by humans at great effort and time costs.

## Environmental Sustainability

The environmental sustainability perspective taken in this report is focused on CO2 emissions. We observe that the computers (currently) produce as much emissions accounted for by human activity as the aviation industry (about 2% each). While acknowledging that the use of computers will only grow, we set out to explore how can the SPL community **help to reduce the growth of the CPU use cycles**? Several areas where SPL helps here are via:

- Removing the need to develop each system from scratch, by re-using common and variable assets of SPL;
- Increasing quality of software, thus reducing cost of testing, and re-work;
- Increasing productivity of developers, whereby a given project requires less CPU cycles to complete;
- Reducing time for bringing systems to market, which in many cases would displace a more environmentally affective practice (e.g., using email instead of a postal service for sending physical letters through a truck fleet, or simulating dangerous processes, such as explorations/explosions, instead of carrying these out in physical world).

In summary to be sustainable at SPL/SBSE: *Define it, measure it, optimize it.*

## 3. SUSTAINABILITY DESING PRINCIPLES AND COMMITMENTS

The Karlskrona manifesto for sustainability design [1] has defined a number of principles and commitments, which, if applied, will promote and support sustainability in and through software systems. Relevance and implications of each of these principles, as perceived by the SPLC 2014 panel on Sustainability in SPL, are discussed below.

*Principle 1: Sustainability transcends multiple disciplines*

*Commitment: Working in sustainability means working with people from across many disciplines, addressing the challenges from multiple perspectives*

Involving multiple disciplines is always key (e.g., projects with SVAI required software engineers, hardware specialists, metallurgists, sales, marketing, management, etc.). Only when involving all relevant people, sustainability can be achieved, e.g., by changing organisational culture to SPL thinking.

However, it has to be noted that this principle is not unique for addressing sustainability concerns, but is inherent for any "big problem".

*Principle 2: Sustainability requires long-term thinking*

*Commitment: Consider multiple timescales, including the long term. Include longer-term indicators in assessment and decisions* Similar to principle 1, this principle also is not unique to addressing sustainability concerns, but is inherent for any "big problem". Yet, this principle is applicable to all scopes of sustainability: relating to both environmental, economic, and social concerns.

The SPL business model supports a "longer-term" thinking than the traditional development models. In particular, through continuous monitoring and adjustment of the current situation, longer-term goals can be better managed. However, this "longer-term" horizon covers only a few months time ahead. This certainly is not sufficiently long-term for addressing the big issues related to environmental and social sustainability.

*Principle 3: It is possible to meet the needs of future generations without sacrificing the prosperity of the current generation.*

*Commitment: Innovation in sustainability can play out as decoupling present and future needs. By moving away from the language of conflict and the trade-off mind-set, we can identify and enact choices that benefit both present and future.*

Innovation is essential (however, re-invention of the wheel is a constant danger and innovation ideas should always be carefully checked with existing solutions). In ideal world each problem will have a solution where (innovation-based) decoupling could resolve trade offs.

However, where such solutions are not found, we advocate adoption of the search-based software-engineering perspective, whereby this principle can be re-formulated as an optimization problem (e.g., finding optimal point between sustainability concerns and those of costs, delivery time, etc.) in a multi-objective trade-off space: "minimal sacrifice for maximal sustainability". Such formulation leads to identification of the Pareto fronts, i.e., a set of (equally "good") solutions whereby interchange increments in various objectives (e.g., sustainability vs. cost) is optimized.

*Principle 4: Sustainability is systemic*

*Commitment: Sustainability is never an isolated property. Systems thinking has to be the starting point for the trans disciplinary common ground of sustainability.*

Similar to principle 1, this principle also is not unique to addressing sustainability concerns, but is inherent for most "big problems".

Yet, we agree that systems thinking is essential in establishing a sustainable SPL-lead business, though may not even be sufficient - as often systems of systems thinking is required. Since a number of systems influence each other (e.g., the payment/promotions system affects the way that people are assigned to roles/teams and so affects people mobility and, consequently, the SPL process and product) this has to be taken into account (which is accounted for in Principle 7).

*Principle 5: Sustainability requires action on multiple levels*

*Commitment: Seek interventions that have the most leverage on a system and consider the opportunity costs: Whenever you are taking*

*action towards sustainability, consider whether this is the most effective way of intervening in comparison to alternative actions (leverage points).*

This principle, again, is not unique to addressing sustainability concerns, but is inherent for most "big problems".

With respect to SPL, we should point out the importance of monitoring and re-evaluation: having captured and compared different alternatives with each other, one is more prepared for selection of the most effective action. This, of course, re-iterates the need for good automation in support for SPL, both on product and context side. In PL we can set up monitoring environment not only to validate configurations, manage variability/commonality and other product or process related data, but also set up (e.g., a machine-learning) environment to collect and interpret economics and analytics-related data.

Another point relevant to leverage points is consideration of the role of the government in IT sector: it has to create a regulatory framework which motivates sustainability requirements. These requirements will, in turn, be propagated by customers to the SPL-based software providers.

*Principle 6: Sustainability applies to both a system and its wider context*
*Commitment: There are at least two spheres to consider in system design:*

- *the sustainability of the system itself:*
- *how it affects overall sustainability of the wider system of which it will be part of*

*It is often useful to distinguish multiple systems during design, to consider the impact on each of them, such as the designed system vs. its production system and usage systems.*

From the search-based software engineering perspective, this principle can be refined into 3 alternative formulations:

a) If the wider context is the *other systems*, we must co-evolve the given system and its environment. This can be formulated as a co-evolution problem for fitness function with two evolving populations whose fitness functions depend on each other

b) if the wider context means different "demanding contexts" (such as people), we need to test systems to check how robust/sustainable they are in the different demanding contexts. So will search for demanding contexts to test the evolving systems.

c) Finally, we can aim to adapt a system to its wider context. One applications of SBSE, can be by using it to search over history of patterns of power consumption and find optimization strategies for adapting the device setups such that power consumption is minimized (e.g., overnight use SBSE to search over the phone use patterns during the past day to optimize its settings for minimal battery power consumption).

*Principle 7: System visibility is a necessary precondition and enabler for sustainability design*

*Commitment: Strive to make the status of the system and its context visible at different levels of abstraction and perspectives to enable participation and informed responsible choice.*

Indeed, visibility is a necessary pre-condition of a good decision-making. Visibility is interpreted as measurability. To paraphrase Kelvin, we can't control what we cannot measure, and we can't affect or optimize what we can't measure.

**Conclusion**

Sustainability is becoming and increasingly important subject in the field of software development and its relevance to and influence on Software Product Lines cannot be denied. It is clear that enomic sustainability is a major driver for applying SPLE within a company, but our study highlights that it is interwoven with all aspects of sustainability in a meriad of ways. For example the ability to scale up development without overstretching the workforce is a direct relation of SPLE to social sustainability.

But while we have noted a strong relation between various aspects of sustainability and the benefits and drawbacks of Software Product Line based development, we also observe the limited and informal understanding of how these relations can be considered and exploited. The strongest understanding exists in the area of economic sustainability where metrics and links to business processes are being explored to inform the effectivenes and benefits of the SPL to the company. Far less of an understanding exists about the relation between Software Product Lines and for example environmental and social sustainability. These fall outside the traditional considerations for SPLE and, as such, have received less attention from industry and the research community.

We call upon SPL community to explore in detail the relations between SPLE and Sustainability to achieve a better understanding of its implications, benefits and drawbacks. From this panel it is clear that SPLE can and does contribute to sustainability, and this positive conurbation must be further maximised.